\documentclass[structabstract]{aa}
\usepackage{graphicx}
\usepackage[varg]{txfonts}
\usepackage{natbib}
\usepackage{color}
\begin{document}

\title{LX Cygni: A carbon star is born\thanks{Based on observations made with the Mercator Telescope, operated on the island of La Palma by the Flemish Community, at the Spanish Observatorio del Roque de los Muchachos of the Instituto de Astrof\'{i}sica de Canarias, and on observations made with the Nordic Optical Telescope, operated by the Nordic Optical Telescope Scientific Association at the Observatorio del Roque de los Muchachos, La Palma, Spain, of the Instituto de Astrofisica de Canarias.}}
\author{S.\ Uttenthaler\inst{1}
  \and
  S.\ Meingast\inst{1}
  \and
  T.\ Lebzelter\inst{1}
  \and 
  B.\ Aringer\inst{2,1}
  \and
  R.\ R.\ Joyce\inst{3}
  \and
  K.\ Hinkle\inst{3}
  \and
  L.\ Guzman-Ramirez\inst{4}
  \and
  R.\ Greimel\inst{5}
}

\institute{University of Vienna, Department of Astrophysics,
  T\"urkenschanzstra\ss e 17, A-1180 Vienna\\
  \email{stefan.uttenthaler@univie.ac.at}
  \and
  Department of Physics and Astronomy G.\ Galilei, University of Padova, Vicolo
  dell'Osservatorio 3, I-35122 Padova, Italy
  \and
  National Optical Astronomy Observatory, P.O.\ Box 26732, Tucson, AZ 85726, USA
  \and
  European Southern Observatory, Alonso de Cord\'ova 3107, Vitacura, Santiago,
  Chile
  \and
  IGAM, Institut f\"ur Physik, Universit\"at Graz, Universit\"atsplatz 5/II,
  8010 Graz, Austria;
  \email{rgreimel@gmail.com}
}

\date{Received May 28, 2015; accepted November 03, 2015}

\abstract
    {The Mira variable LX Cygni (LX~Cyg) has shown a dramatic increase of its
      pulsation period in the recent decades and is appearing to undergo an important
      transition in its evolution.}
    {We aim to investigate the spectral type evolution of this star over       recent decades as well as during one pulsation cycle in more detail and
      discuss it in connection with the period evolution.}
    {We present optical, near- and mid-infrared low-resolution as well as
      optical high-resolution spectra to determine the current spectral type.
      The optical spectrum of LX~Cyg has been followed for more than one
      pulsation cycle. We compare recent spectra  to archival spectra to trace
      the spectral type evolution, and we analyse a Spitzer mid-IR spectrum  for
      the presence of molecular and dust features. Furthermore, the current
      pulsation period is derived from AAVSO data.}
    {We found that the spectral type of LX~Cyg changed from S to C sometime
      between 1975 and 2008. Currently, the spectral type C is stable during a
      pulsation cycle. We show that spectral features typical of C-type
      stars are present in its spectrum from $\sim0.5$ to 14\,$\mu{\rm{m    }}$, and attribute an
      emission feature at 10.7\,$\mu{\rm{m}}$  to SiC grains. Within only 20 years, the
      pulsation period of LX~Cyg has increased from $\sim460$\,d to $\sim580$\,d
       and is stable now.}
    {We conclude that the change in spectral type and  increase in pulsation
      period happened simultaneously and are causally connected. Both a recent
      thermal pulse (TP) and a simple surface temperature decrease appear
      unlikely to explain the observations. We therefore suggest that the
      underlying mechanism is related to a recent third dredge-up mixing event
      that brought up carbon from the interior of the star, i.e.\ that a genuine
      abundance change happened. We propose that LX~Cyg is a rare transition
      type object that is uniquely suited to study the transformation from
      oxygen- to carbon-rich stars in detail.}


\maketitle


\section{Introduction}\label{Intro} 

Mira variables are red giant stars thought to be in the asymptotic giant branch
(AGB) phase of evolution. In this phase, important changes happen in the star
due to the addition of nucleosynthesis products to its atmosphere by a deep
mixing process called the third dredge-up (3DUP). Most notably, carbon
($^{12}$C) is added, which has a major impact on the molecular equilibrium in
the cool outer layers of the star. The crucial parameter determining the
appearance of the star's spectrum is the ratio of the number of oxygen to carbon
atoms. Most of O and C are locked up in carbon monoxide (CO), and the more
abundant of these elements forms other molecules that dominate the spectrum.
Metal oxides (TiO, VO), in the case of an excess of oxygen, form the
spectral bands typical of M-type stars, whereas carbon-bearing molecules (CN,
C$_2$, C$_2$H$_2$) form in carbon-rich stars of type C. The star is of
intermediate spectral type (MS, S, SC) as the C/O ratio increases towards unity;
see \citet{GC09} for details of spectral classification of late-type giants.
Thus, the sequence M -- MS -- S -- SC -- C is thought to reflect an evolutionary
path that a star takes on the AGB. It is known that due to the pulsation, Mira
variables change their spectral subtype within the main spectral types because
of temperature variations along a pulsation cycle \citep{Ter69}. True spectral
type changes are rare. Only a few examples have been claimed, and one of those is
\object{BH Cru} \citep{Whi99}. This star seems to have changed its spectral type
from SC to C \citep{TLE85} with little cyclical variation after the change
(Lloyd Evans, private communication). Alternative explanations have been put
forward that do not require a genuine chemical abundance change \citep{Zij04}.

Also, most Mira variables have pulsation periods that are more or less stable
over decades to centuries. \citet{TMW05} investigated light curves of 547 Miras
with a long baseline of visual observations. Eight Miras were found to have
changed their pulsation periods systematically at the significance level of
$6\sigma$ or greater, either increasing or decreasing, in the past decades.
Further candidates for stars with changing pulsation periods were presented by
\citet{LA11}. The mechanism usually put forward to explain the changing
pulsation period is the violent ignition of the dormant He-burning shell during
a so-called thermal pulse (TP) or He-shell flash \citep{Wood75}. The radius of
the star changes in reaction to the changed energy supply (surface luminosity)
during the TP, which in turn affects the pulsation period \citep{VW93}. In the
aftermath of a TP, a 3DUP event may occur when the convective envelope deepens
to reach stellar layers where nuclear processing has taken place before. In this
way, a 3DUP event is intimately connected to a TP.

The star with the strongest period increase in the sample of \citet{TMW05} is
\object{LX Cygni} (LX~Cyg). This star is very similar to BH~Cru in that both underwent
a large period increase and were classified as S/SC type before the period
increase. After its period increase, BH~Cru was observed to be a C-type Mira
\citep{Whi99}. In this paper, we report spectral observations of LX~Cyg to
analyse its current spectral type and spectral type evolution during recent
decades. We demonstrate that  with respect to the spectral type after the
period increase both BH~Cru and LX~Cyg also show the same behaviour, which is
probably the result of the same underlying physical processes.

This paper is structured in the following way. In Sect.~\ref{Obs} the
observations are presented. Section~\ref{analys} first analyses the period
evolution of LX~Cyg (Sect.~\ref{Period_evol}) before continuing with the
spectral analysis, both for the current spectral type (Sect.~\ref{current_type})
as well as its spectral type evolution (Sect.~\ref{SpT_change}). The results are
discussed in Sect.~\ref{disc}, and conclusions are drawn in
Sect.~\ref{conclusio}.

\section{Observations}\label{Obs}

We based this study on the experience with BH~Cru \citep{Whi99,Utt11}, a Mira
that exhibited a period increase similar to that of LX~Cyg along with a change
of spectral type from S/SC to C. As a starting point of this study, we obtained
a high-resolution optical spectrum of LX~Cyg to investigate its current spectral
type. The spectrum was obtained with the Hermes fibre-fed spectrograph
\citep{Ras11} at the 1.2\,m Mercator telescope on the island of La Palma, Spain,
in high-resolution mode ($R=85\,000$). Two exposures of 1800\,s integration time
each were taken on 24 August 2011, covering the wavelength range 377 -- 900\,nm.
However, shortwards of 500\,nm the flux is essentially zero in these spectra.
This spectrum is presented and discussed in Fig.~\ref{Barnbaum}; see
Sect.~\ref{current_type}.

Because of claims that the change in spectral type from SC to C could be due to
a temperature decrease in the atmosphere rather than a true change in chemical
composition \citep{Zij04}, we monitored LX~Cyg with low-resolution optical
spectroscopy over about one pulsation cycle, i.e.\ between 28 November 2011 and
13 August 2013. The aim of these observations was to see if the spectral type
depends on the pulsation phase. The observations were obtained with the 0.8\,m
Cassegrain telescope of the Vienna University Observatory (VUO) and a DSS-7
spectrograph manufactured by SBIG. The DSS-7 grating spectrograph provides for
a maximum dispersion of $\sim0.55$\,nm/pixel, corresponding to a spectral
resolving power of $R\approx500$.
With this camera, the wavelength coverage is approximately $480-820$\,nm. At
each observing date, several exposures of LX~Cyg of 300\,s integration time were
obtained, but here only the individual spectra with the highest signal-to-noise
ratio are presented. Care was taken to position the slit such that no nearby
star contaminated the spectrum of the target. Other than LX~Cyg, known SC-type
stars such as \object{FU Mon}, \object{CY Cyg}, \object{GP Ori}, and
\object{R CMi,} were obtained with the same instrument for comparison.
These data are discussed in more detail in Sects. \ref{current_type} and
\ref{spec_mon}.

%

On 21 October 2012, we also obtained $R=2500$ spectra with NOTCam \citep{Abb00}
at the Nordic Optical Telescope on the island of La Palma, Spain, covering the
range $\sim 820 - 2360$\,nm. These observations where motivated by the fact that
LX~Cyg was observed by \citet{Joy98} at low resolution in the J-band in 1994;
our spectra provide for a check of the spectral changes that might have
happened between 1994 and 2012. The data are also useful to compare with other
observations from the literature. See Sects. \ref{current_type} and
\ref{SpT_change} for a discussion of these data.
Table~\ref{own_obs} lists all the important information of our spectral
observations.

\begin{table}
\caption{Data of our spectral observations}
\label{own_obs}\centering
\begin{tabular}{llrr}
\hline\hline
Date       & Instrument      & Res.\ power           & Wavelength \\
dd/mm/yyyy &                 & $R=\lambda/\Delta\lambda$ & (nm)   \\
\hline
24/08/2011 & Hermes/Mercator & 85\,000 & $\sim377-900$ \\
28/11/2011 & DSS-7/VUO       &     500 & $\sim450-858$ \\
20/12/2011 & DSS-7/VUO       &     500 & $\sim450-856$ \\
18/01/2012 & DSS-7/VUO       &     500 & $\sim450-858$ \\
15/06/2012 & DSS-7/VUO       &     500 & $\sim402-815$ \\
23/07/2012 & DSS-7/VUO       &     500 & $\sim400-806$ \\
14/08/2012 & DSS-7/VUO       &     500 & $\sim405-815$ \\
11/09/2012 & DSS-7/VUO       &     500 & $\sim403-816$ \\
21/10/2012 & NOTCam/NOT      &  2\,500 & $\sim830-2360$\\
16/11/2012 & DSS-7/VUO       &     500 & $\sim408-820$ \\
12/12/2012 & DSS-7/VUO       &     500 & $\sim408-820$ \\
13/08/2013 & DSS-7/VUO       &     500 & $\sim410-820$ \\
\hline
\end{tabular}\\
\end{table}

\begin{figure}
  \centering
  \includegraphics[width=\columnwidth,bb=82 371 570 603, clip]{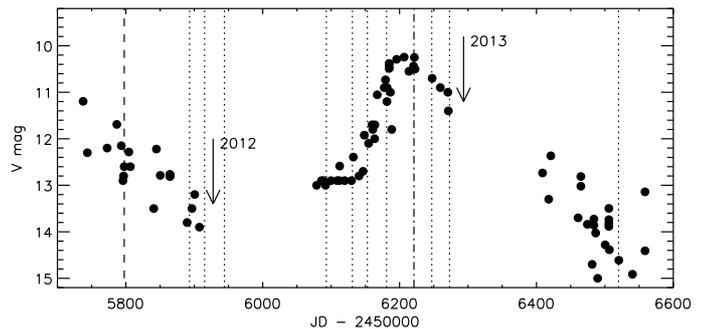}
  \caption{Visual and V-band photometric mean AAVSO light curve of LX~Cyg
    between JD 2\,455\,700 and JD 2\,456\,600. The dashed line indicates the time of
    the Hermes/Mercator high-resolution observations, the dash-dotted line the
    NOTCam near-IR observations, and the dotted lines the optical low-resolution
    observations at the Vienna University Observatory. Arrows denote 1 January
    2012 and 2013.
  }
  \label{LXCygLC}
\end{figure}

Figure~\ref{LXCygLC} shows the AAVSO\footnote{{\tt http://www.aavso.org/}}
visual light curve of LX~Cyg, for the time span in which we obtained spectral
observations, to indicate the visual phase. The data points are 10-day means of
visual and V-band photometric observations. There is no systematic shift between
the visual estimates and V-band observations, so they were averaged, as
either of them alone would  trace the light curve insufficiently. 
In total, the spectral observations span more than one pulsation cycle.

Finally, we also searched the {\it Spitzer} space telescope archive for
observations of LX~Cyg. Indeed, observations were carried out on 05 December 2008 in
programme ID 50717, AOR: 27668480 (PI: M.\ Creech-Eakman) with the Infrared
Spectrograph \citep[IRS][]{Hou04}. The spectrum covers the range between 5.2
and 37.2\,$\mu m$ by using the short-low (SL: $5.2-14.5$\,$\mu{\rm{m}}$;
$64<R<128$), short-high (SH: $9.9-19.6$\,$\mu{\rm{m}}$; $R\sim600$) and
long-high (LH: $18.7-37.2$\,$\mu{\rm{m}}$; $R\sim600$) modules. The data were
retrieved from the {\it Spitzer} Science Center Data Archive using Leopard. The
{\it Spitzer} IRS Custom Extractor (SPICE) was used to perform the extraction of the
spectra for each nod position from the 2D images. The spectra were cleaned for
bad data points, spurious jumps, and glitches, and then they were combined and
merged to create a single $5-37$\,$\mu{\rm{m}}$ spectrum. We did not correct for
flux offsets between different wavelengths ranges, but these are minor. The
{\it Spitzer} spectrum allows for an investigation of the type of dust forming
around the star; this will be discussed in more depth in Sect.~\ref{sect_CSE}
(Figs.~\ref{fig_SED} and \ref{Spitzer}).

\section{Analysis and results}\label{analys}

\subsection{Period evolution}\label{Period_evol}

The variability of LX~Cyg was first pointed out by \citet{Hof30}. The dramatic period
increase of this star was first reported by \citet{TMP03} and confirmed by
\citet{BA03} as well as \citet{Zij04}. \citet{TMP03} analysed the AAVSO
observations that commenced in 1967 and find a period increase from $\sim460$ to
$\sim580$\,d. The period increase seems to have started in 1975; the strongest
growth happened between 1983 and 1995 when the period increased at a rate of
$\sim7.6$\,d/yr. All observations prior to 1967 indicate that the period was
close to $\sim460$\,d, suggesting that it was relatively stable, apart from
cycle-to-cycle variations that are seen in many Miras of such long period.

We used AAVSO visual data to determine the pulsation period of LX~Cyg between JD
2\,452\,000 -- 2\,456\,626 (31 March 2001 -- 30 November 2013), including eight
maxima. The data were analysed with the programme {\tt period04} \citep{LB05}. A
mean period of 588.3\,d was found from these data. It appears that the period
has not significantly increased since 1995. We therefore conclude that the
pulsation period of LX~Cyg was stable at $\sim460$\,d until 1975, followed by a
marked period increase of nearly 25\% up to 1995, when it stabilised at
$\sim580$\,d. This means that the expansion of the star that caused the period
increase has stopped by now. BH~Cru showed a very similar evolution of pulsation
period, although it was claimed by \citet{Zij04} that it may have had a long
pulsation period a few decades before the recently observed period increase, and
that its period may be intrinsically unstable.

\subsection{Current spectral type}\label{current_type}

Several comparisons were done to check the current spectral type of LX~Cyg to
classify it among known SC- and C-type stars, and to check for any spectral
differences.

\begin{figure}
  \centering
  \includegraphics[width=\columnwidth,bb=92 372 532 693, clip]{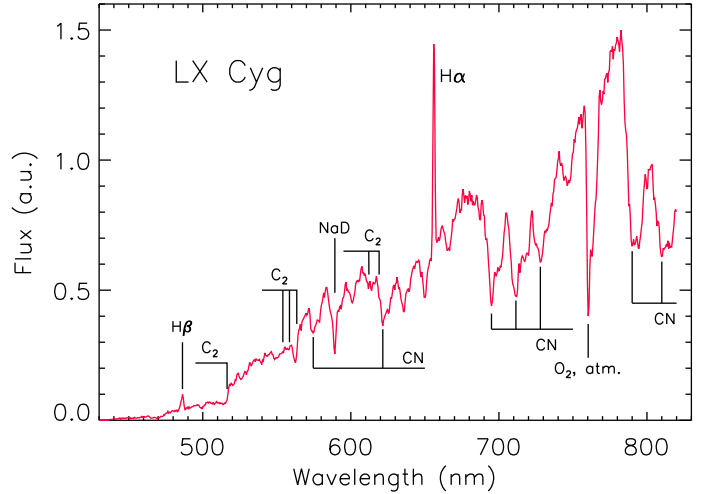}
  \caption{Low-resolution optical spectrum of LX~Cyg on 16 November 2012 with
    the main spectral features labelled.}
  \label{screen}
\end{figure}

In the optical spectral range, we followed the SC spectral-type criteria of
\citet{KB80}, which are reproduced in Table~\ref{SC_criteria}. Bands of ZrO are
present in types SC\,X/7, which equals SX/7, or earlier abundance types. Here,
``X'' refers to the temperature type. A high signal-to-noise ratio spectrum of
LX~Cyg obtained with the DSS-7 spectrograph at the Vienna University Observatory
gives a good overview of the molecular features present in the star as well as
the general flux distribution throughout the optical spectral range. The
spectrum obtained on 16 November 2012 as part of our spectral monitoring
campaign (Sect.~\ref{spec_mon}) is shown in Fig.~\ref{screen}. On that date,
LX~Cyg was close to maximum visual light (Fig.~\ref{LXCygLC}). The main spectral
features in that spectrum are labelled. Prominent bands of ZrO would be located
at 555.2, 584.9, 613.6, 647.4, and 649.5\,nm \citep{GC09}, but they are absent.
Bands of CN and C$_2$ dominate the spectrum. The spectra of known SC stars
obtained with the same instrumentation indicates that, if at all, none of these shows
such prominent C$_2$ bands as LX~Cyg. We conclude that LX~Cyg cannot be of type
SX/7 or earlier abundance type.

\begin{table}
\caption{Classification criteria for the SC spectral types, reproduced from
  \citet{KB80}.}
\label{SC_criteria}\centering
\begin{tabular}{llc}
\hline\hline
Spectral type     & Criteria for C/O                      & Estimated \\
                  &                                       & C/O \\
\hline
SX/7 = SC\,X/7    & ZrO weaker. D lines strong.           & 0.99 \\
SC\,X/8           & No ZrO or C$_2$. D lines very strong. & 1.00 \\
SC\,X/9           & C$_2$ very weak. D lines very strong. & 1.02 \\
SC\,X/10 = C\,X,2 & C$_2$ weak. D lines strong.           & 1.1: \\
\hline
\end{tabular}\\
\tablefoot{X denotes the temperature type; D lines are Na\,D lines.}
\end{table}

\begin{figure}
  \centering
  \includegraphics[width=\columnwidth,bb=78 369 572 569, clip]{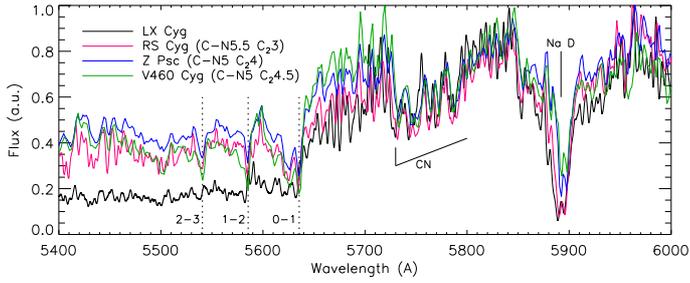}
  \caption{Comparison between the N-type spectral standard stars RS~Cyg
  (C-N5.5 C$_2$3, red), Z~Psc (C-N5 C$_2$4, blue), and V460~Cyg (C-N5 C$_2$4.5,
  green) from \citet{Bar96} with the Hermes spectrum of LX~Cyg smoothed to the
  resolution of the atlas spectra ($R=2500$, black). The spectra are scaled to
  have the same maximum flux in the shown wavelength range. Vertical dotted
  lines indicate the wavelengths of C$_2$ band heads from the Swan system
  \citep{Kip04}.}\label{Barnbaum}
\end{figure}

The best distinguishing feature between the SC subtypes is the strength of the
C$_2$ bands. To estimate their strength, we compared the Hermes spectrum to the
spectral atlas of carbon stars of \citet{Bar96}. The Hermes spectrum was
smoothed to the resolution of the atlas spectra ($R=2500$) for this comparison.
In particular, the spectrum was compared to the N-type carbon stars
\object{RS Cyg}, \object{Z Psc}, and \object{V460 Cyg,} which are presented by
\citet{Bar96} to illustrate the carbon abundance scale of N-type stars; see
their Fig.~1c. These three stars have C$_2$ indices of 3, 4, and 4.5,
respectively. This comparison is shown in Fig.~\ref{Barnbaum}. Dotted vertical
lines in that figure indicate the wavelengths of the $0-1$, $1-2$, and $2-3$
band heads of the C$_2$ Swan system that characterise typical carbon stars
\citep{Kip04}. The C$_2$ bands are not weak in LX~Cyg. They are clearly
discernible, indicating a spectral subtype of at least SC\,X/10, which equals
C\,X,2 on the carbon star C/O abundance scale, or even higher abundance index.
From this, one would currently not classify it as SC-type star.

A difference between LX~Cyg and the atlas spectra of \citet{Bar96} is that
LX~Cyg seems to faint more strongly at wavelengths shorter than $\sim5700$\,\AA\
than the atlas stars do. This may be caused by a higher amount of dust around
LX~Cyg than around the other stars. This is confirmed by the 2MASS $J-K$ colour
of the stars. While RS~Cyg, the reddest of the three atlas stars, has 1.885 in
this colour, LX~Cyg has $J-K_{\rm S}=2.199$. The other two stars are even bluer
at $J-K_{\rm S}\approx1.52$. As we show below (Sect.~\ref{sect_CSE}), the star's
mid-IR spectrum  also favours the presence of dust around LX~Cyg. Among these
four stars, the Na\,D line is indeed strongest in LX~Cyg. The comparison is
complicated by the fact that LX~Cyg is a Mira variable with a visual amplitude
of $\sim4$\,mag, whereas the spectral atlas stars are much less variable
(semi-regular variables of type SRa and SRb).

The high-resolution Hermes spectrum was also searched for lines of heavy
elements that are formed during the slow neutron-capture (s-)process in the
interior of AGB stars. All the heavy-element lines that also \citet{AW98}
detected in their high-resolution spectrum of BH~Cru could be detected.
Therefore, we can also conclude that LX~Cyg is enriched in s-process elements
and that it is a true TP-AGB star. We caution that the Hermes spectrum of LX~Cyg
is very complex, hence quantitative measurements are unfortunately difficult to
make.

\begin{figure}
  \centering
  \includegraphics[width=\columnwidth,bb=83 369 536 698, clip]{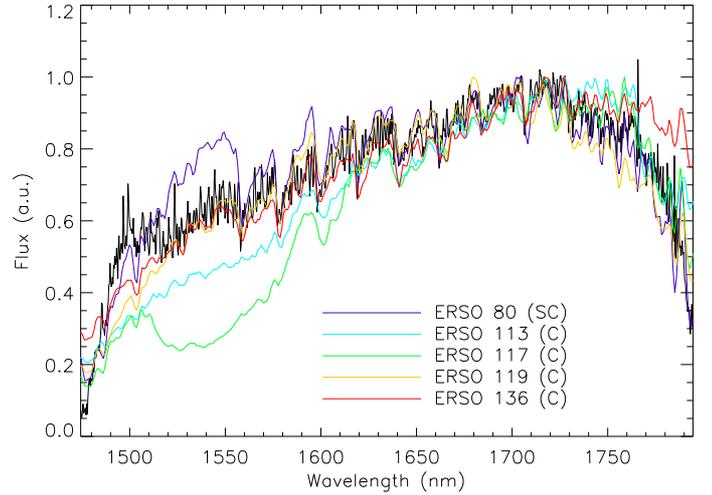}
  \caption{NOTCam H-band spectrum of LX~Cyg (black line) compared to four C-type
    and one SC-type star identified by \citet{Wri09}; see legend.}
  \label{Wri09_comp}
\end{figure}

At longer wavelengths, the NOTCam spectrum of LX~Cyg was compared to spectra of
C- and SC-type stars identified by \citet{Wri09}. Figure~\ref{Wri09_comp} shows
this comparison for the H band, which contains a feature at $\sim1530$\,nm that
is probably due to C$_2$H$_2$. It appears that SC- and C-type stars differ
markedly in this feature. The one SC-type star in the sample of \citet{Wri09},
the extremely red stellar object (ERSO) no.~80, shows a lack of opacity at the
wavelength of this feature, compared to a flux depression in the C-type stars.
The difference between the stars in the strength of this feature may be due to a
difference in temperature (the feature becoming stronger at lower temperature)
and/or due to a difference in composition (becoming stronger at higher C/O).
Figure~\ref{Wri09_comp} shows that LX~Cyg is very similar to two of the C-type
stars, namely ERSO~119 and ERSO~136. Besides this C$_2$H$_2$ feature, also the
CO $\Delta\nu=3$ bands at $\sim1558$\,nm, $\sim1577$\,nm, $\sim1598$\,nm, etc.,
are clearly visible in all stars shown in Fig.~\ref{Wri09_comp}. From this
comparison, we can conclude that LX~Cyg is now a C-type star.

\subsection{Circumstellar environment of LX~Cyg}\label{sect_CSE}

\begin{figure}
  \centering
  \includegraphics[width=\columnwidth,bb=84 369 545 699, clip]{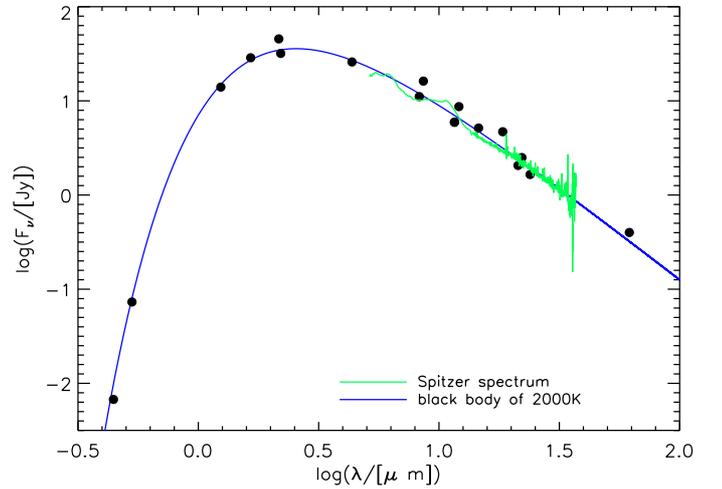}
  \caption{Observed SED of LX~Cyg (black dots). The {\it Spitzer} IRS spectrum
    is plotted in green. The blue curve is a black-body spectrum of 2000\,K,
    scaled to have the same median flux as the {\it Spitzer} spectrum in the
    range $20-24$\,$\mu{\rm m}$.}
  \label{fig_SED}
\end{figure}

To help analyse the {\it Spitzer} IRS spectrum of LX~Cyg, we first inspect its
spectral energy distribution (SED); see Fig.~\ref{fig_SED}. Photometry was
collected from VizieR\footnote{\tt http://vizier.u-strasbg.fr/viz-bin/VizieR},
except for the fluxes in the B and V bands, for which cycle-averaged fluxes from
AAVSO data have been used. The wavelength range from the B-band
(0.444\,$\mu{\rm m}$) out to the IRAS~60\,$\mu{\rm m}$ band is covered by the
SED. A black-body spectrum of a temperature of 2000\,K is included in that figure
(blue line). This spectrum is scaled to have the same median flux as the {\it Spitzer}
spectrum (green line) in the range $20-24$\,$\mu{\rm m}$. The SED of LX~Cyg is
well represented by the 2000\,K black-body spectrum.

\begin{figure}
  \centering
  \includegraphics[width=\columnwidth,bb=87 369 537 699, clip]{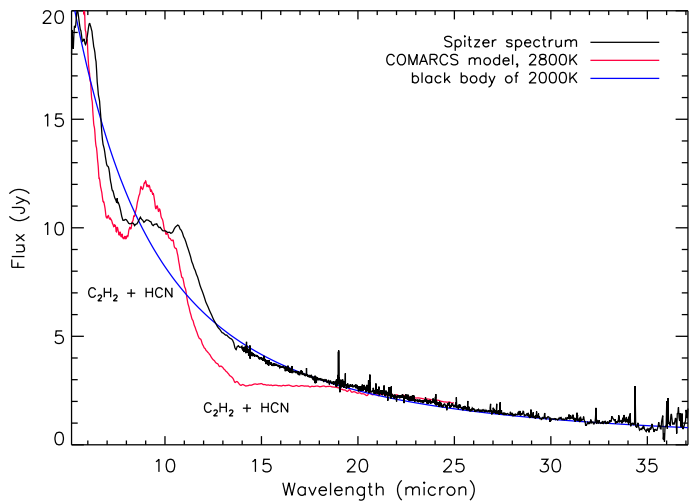}
  \caption{{\it Spitzer} IRS spectrum of LX~Cyg (black line). Over-plotted
      in blue is the same black-body spectrum as in Fig.~\ref{fig_SED}, as well
      as a scaled dust-free C-star model spectrum from
      \citet[][red line]{Ari09}.}
  \label{Spitzer}
\end{figure}

The {\it Spitzer} spectrum is shown in linear scale in more detail in
Fig.~\ref{Spitzer}. The black-body spectrum, scaled in the same way as in
Fig.~\ref{fig_SED}, is an excellent fit to the spectrum at wavelengths longer
than $\sim17$\,$\mu{\rm m}$. No dust features are visible between 17 and 37
micron. At shorter wavelengths, there are regions with an excess or a deficit of
flux compared to the black body. We interpret these features with the help of
an over-plotted COMARCS model spectrum of a dust-free carbon star from
\citet[][red line]{Ari09}. This model has the following stellar parameters:
$T_{\rm eff}=2800$\,K, $\log g\,[{\rm cm\,s^{-2}}]= 0.0$, $M=1.0M_{\sun}$,
$Z=1.0Z_{\sun}$, micro-turbulence $\xi=2.5$\,km\,s$^{-1}$, and C/O=1.05. The model
spectrum is scaled in the same way as the {\it Spitzer} and black-body spectra.
The surplus flux at $\sim9.5$\,$\mu{\rm m}$ compared to the black body is not an
emission component, but rather is a result of the scaling to the
pseudo-continuum, not the true continuum. With this COMARCS model, we identify
the flux depressions at $\sim7$\,$\mu{\rm m}$ and $\sim14$\,$\mu{\rm m}$ with
blended molecular bands of C$_2$H$_2$ $+$ HCN, typical of carbon stars. However,
we can clearly see two main emission features peaking at $\sim6.2$\,$\mu{\rm m}$
and $\sim10.7$\,$\mu{\rm m}$. The latter agrees with a peak from SiC dust grain
emission predicted by \citet{Mut99}, which are expected to form in carbon-rich
environments. We therefore assume that this emission peak stems from SiC grains.


The peak at 6.2\,$\mu{\rm m}$ is more uncertain to interpret. It coincides with
an emission feature known from polycyclic aromatic hydrocarbon (PAH) molecules,
which are also expected to be formed in C-rich environments such as atmospheres
of carbon stars \citep{All89}. However, there are only a few AGB stars where
mid-IR features of PAHs have been detected; see \citet{Smo10} for a discussion.
Also, the 6.2\,$\mu{\rm m}$ feature may be expected to be accompanied by other
PAH features at 7.9, 8.6, and 11.2\,$\mu{\rm m}$, which we cannot detect. It is
unclear if the conditions can be such that only the 6.2\,$\mu{\rm m}$ feature
emits. As PAH features have been detected in S- as well as in C-type stars
\citep[][and references therein]{Smo10}, there is not a strong discriminant between
those spectral types. Non-equilibrium, shock-driven chemistry may be responsible
for PAH formation via C$_2$H$_2$ at C/O$\sim$1.0 \citep{Cher92}. Alternatively,
the feature could be the result of an absorption band at 5.5\,$\mu{\rm m}$ by
C$_3$ \citep{Gau04}. However, this would probably require a somewhat high C/O
ratio, which we consider unlikely.

There are two more arguments for the presence of dust around LX~Cyg. Both have
their physical origin in the fact that short-wavelength photons are absorbed by
dust grains and their energy is re-emitted at longer wavelengths in the mid-IR
range, thereby shifting the maximum flux towards longer wavelengths. The first
argument is the $J-K_{\rm S}$ colour of LX~Cyg. Using 2MASS photometry dated
15 June 2000, $J-K_{\rm S}=2.199$ is found for the star. This is redder than any
of the 12 SC stars observed by \citet{Joy98}, which reach 2.062 at most. It
is also redder than all dust-free COMARCS C-star models of \citet{Ari09}. On the
other hand, the carbon-rich Miras from \citet{Whi06} occupy colours in the range
$\sim1.6$ to 6.0. With the relation between $J-K_{\rm S}$ and the mass-loss rate
given in \citet{Now13}, we derive
$\dot M\approx4\times10^{-7}M_{\sun}{\rm yr}^{-1}$ for LX~Cyg. This mass-loss rate
is relatively low among the C stars. Secondly, the temperature of 2000\,K of the
black-body spectrum required to approximate the SED of LX~Cyg is quite low. Also,
AAVSO visual and V-band observations of the past 15 years suggest that LX~Cyg is
becoming fainter in the optical, which suggests that more dust has recently
formed around the star. This dust may not only be present in the form of SiC
grains as concluded above, but may also be amorphous carbon grains that do not
have any spectral features to identify them.

In summary, the circumstellar environment of LX~Cyg as evidenced by mid-IR
molecular and dust features detected in the {\it Spitzer} IRS spectrum as well
as its SED, suggests a carbon-rich nature of the star.

\subsection{Spectral monitoring}\label{spec_mon}

\begin{figure}
  \centering
  \includegraphics[width=\columnwidth,bb=96 370 458 849, clip]{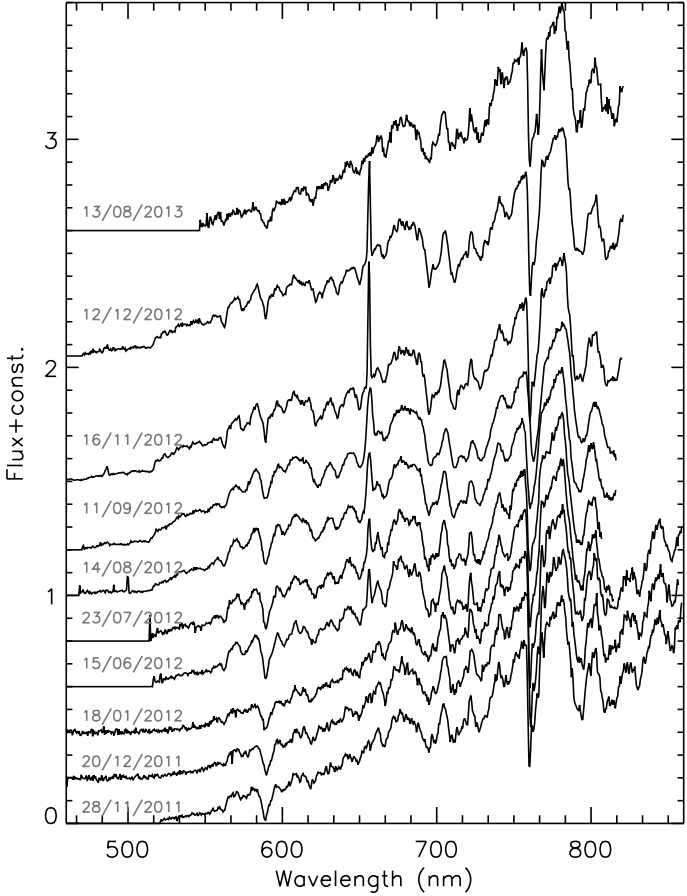}
  \caption{Low-resolution optical spectra of LX Cyg obtained with the DSS-7
    spectrograph at the 0.8\,m Cassegrain telescope of the VUO in chronological
    order, from bottom to top.}
  \label{optlowres}
\end{figure}

In the course of our spectral monitoring programme with the 0.8\,m telescope at
VUO, observations have been obtained on 10 dates between 28 November 2011 and
13 August 2013, spanning 624 days. The spectral evolution in this period is
shown in Fig.~\ref{optlowres}. Even though some of the CN bands are weaker in
the fainter phases of the cycle, this figure clearly demonstrates that
LX~Cyg is of stable spectral type C. The spectrum is dominated by bands of C$_2$
and CN. The C$_2$ bands are strongest at phases around visual maximum. As the
star approaches maximum light, the H$\alpha$ line starts to emit. Shortly after
maximum light (16 November 2012), also the H$\beta$ line is visible in emission;
this spectrum is shown in more detail in Fig.~\ref{screen} with the main
features labelled.

\subsection{When did the change in spectral type happen?}\label{SpT_change}

As has been shown in previous sections, LX~Cyg most probably is a stable
carbon star now. This leads to the question: When did the change in spectral
type happen?

Table~\ref{historic_spt} collects in chronological order all spectral type
determinations of LX~Cyg that we found in the literature. There are at least 13
observations spanning almost 27 years, all of them reporting LX~Cyg with
spectral type S or SC. \citet{Ter69} lists LX~Cyg as showing only bands of LaO
in the red ($680-880$\,nm). \citet{BN57} report LX~Cyg as being of S type with
LaO bands present and being ``very red'', based on the observations published by
\citet{CN56}. These latter authors report that the star has an unidentified
absorption band at 850\,nm, in addition to strong LaO bands at 740 and 790\,nm.
C$_2$ lines are only weakly present at one occasion in 1978 when LX~Cyg was
assigned the abundance index 9 by \citet{KB80}; see Table~\ref{SC_criteria}.

\begin{table}
\caption{Historic spectral type determinations of LX~Cyg.}
\label{historic_spt}\centering
\begin{tabular}{lll}
\hline\hline
Date       & Sp.\ Type     & Reference \\ 
dd/mm/yyyy &               &           \\ 
\hline
22/10/1952 & S\,LaO\,3     & \citet{Ter69} \\ 
05/09/1954 & S\,LaO\,3     & \citet{Ter69} \\ 
20/08/1955 & S             & \citet{CN56}  \\ 
27/09/1956 & S\,LaO\,2-3   & \citet{Ter69} \\ 
05/10/1957 & S\,LaO\,2     & \citet{Ter69} \\ 
12/09/1958 & S\,LaO\,2:    & \citet{Ter69} \\ 
07/06/1962 & S\,LaO\,3     & \citet{Ter69} \\ 
03/10/1965 & S\,LaO\,1     & \citet{Ter69} \\ 
27/08/1975 & S5.5e:\,Zr\,1 & \citet{Ake79} \\ 
14/11/1976 & SC3e\,Zr\,0.5 & \citet{Ake79} \\ 
18/10/1977 & SC8/8.5e      & \citet{KB80}  \\ 
29/10/1978 & SC7/9--e      & \citet{KB80}  \\ 
02/08/1979 & SC4:/8e       & \citet{KB80}  \\ 
\hline
\end{tabular}\\
\end{table}

\begin{figure}
  \centering
  \includegraphics[width=\columnwidth,bb=84 371 544 698, clip]{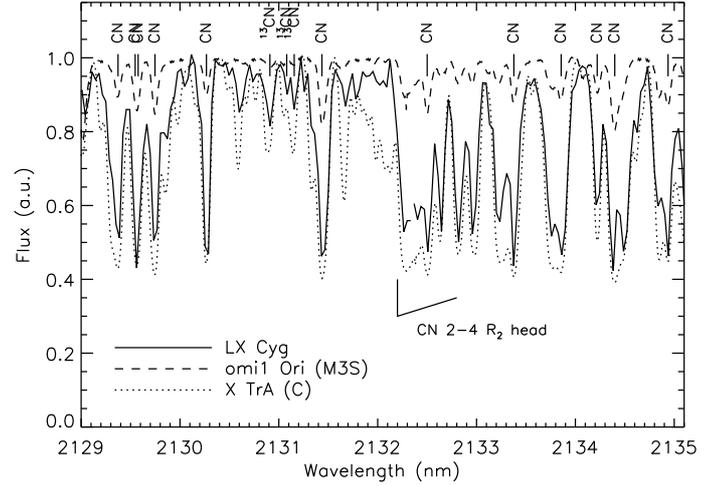}
  \caption{Comparison of the FTS K-band spectrum of LX~Cyg obtained by
    \citet[][solid line]{DW87} with CRIRES-POP \citep{Leb12} spectra of the
    M3S-type giant $o^1$ Ori (dashed line) and the C5 carbon star X~TrA
    (dotted line). $^{12}$CN line identifications from the Arcturus atlas of
    \citet{Hin95}, denoted as CN, as well as $^{13}$CN lines from \citet{Lam86}
    are also shown.}
  \label{Kband}
\end{figure}

After 1979, spectral observations of the star became rare. A high-resolution FTS
spectrum in the K band ($\sim1900-2500$\,nm, $R\approx50\,000$) was obtained on
14 October 1984 (JD 2\,445\,987) by \citet{DW87}. They observe that LX~Cyg is their
``most difficult case'' in a sample of seven S and SC stars. Unfortunately, the
K band does not contain spectral features that are established for spectral
(sub-)type determination of cool giants. Nevertheless, it is instructive to
inspect the strength of the CN lines. We compared the archival FTS spectrum of
LX~Cyg obtained by \citet{DW87} with high-resolution spectra of the M3S-type
giant \object{$o^{1}$ Ori} and the C5,5 carbon star \object{X TrA} obtained with
CRIRES at the VLT in the CRIRES-POP programme \citep{Leb12}; see
Fig.~\ref{Kband}. In that figure, $^{12}$CN as well as a few $^{13}$CN line
identifications from \citet{Hin95} and \citet[][their Fig.~4]{Lam86} are shown,
respectively. \citet{WH96} report that ``CN is the major contributor to
the jumble of lines'' in that spectral region (cf.\ their Fig.~5). The CN lines
are much stronger in LX~Cyg than in $o^1$~Ori, but somewhat weaker than in
X~TrA. The $^{13}$CN lines are weaker in LX~Cyg, which suggests that it has a
lower $^{13}$C abundance than X~TrA.

The same spectrum of LX~Cyg was analysed by \citet{Hin00}, who classified
two-micron infrared spectra of long-period variables into five categories
according to the strength of H$_2$O and CN lines. LX~Cyg falls in their category
``c'', which equals the C spectral type. The authors also point out occasional
surprising differences between the category that they had assigned and the
published spectral type, which is obviously also the case for LX~Cyg. We thus
conclude that it is possible that LX~Cyg was already a carbon star at the time
of observation by \citet{DW87} in 1984.

\begin{figure}
  \centering
  \includegraphics[width=\columnwidth,bb=78 367 561 796, clip]{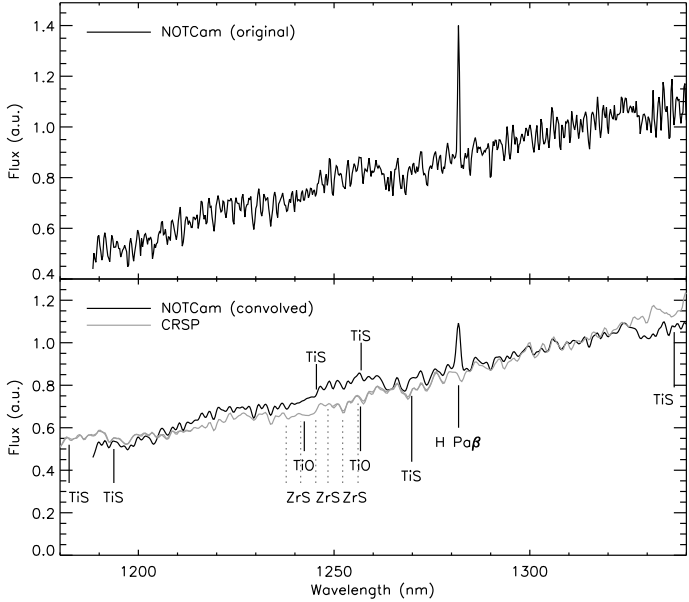}
  \caption{Upper panel: original NOTCam spectrum of LX~Cyg at a resolving
    power of $R=2500$. Lower panel: The CRSP \citep{Joy98} and convolved
    NOTCam J-band spectrum of LX~Cyg. Positions of molecular band heads, as
    identified by \citet{Joy98} to be typically present in S-type stars, are
    indicated. Also the H~Pa\,$\beta$ line at 1282\,nm is indicated, which appears in
    emission in the NOTCam spectrum taken close to maximum visual light.}
  \label{Jband}
\end{figure}

The next spectral observations of LX~Cyg were obtained by \citet{Joy98} on 16
April 1994 (JD 2\,449\,458). These observations were carried out with the Kitt Peak
cryogenic spectrograph (CRSP) in the J band ($\sim1100-1350$\,nm) at a resolving
power of $R=1100$. The CRSP spectrum is compared to the NOTCam spectrum obtained
on 21 October 2012 in the overlapping wavelength range in the lower panel of
Fig.~\ref{Jband}. For better comparison, the NOTCam spectrum was convolved
with a Gaussian kernel to the CRSP resolving power of $R=1100$. The agreement
between the spectra is very good. The features are essentially the same, small
differences such as around $\sim1250$\,nm may be due to the different pulsation
phase at the time of observations or instrumental issues (e.g.\ flat-fielding).
Longwards of 1330\,nm, the slight mismatch between the spectra probably is due to
telluric absorption by H$_2$O. One noteworthy difference between the spectra is
the H Paschen $\beta$ line, which appears in emission in the NOTCam spectrum.
This is in agreement with the fact that other H lines are also in emission
around maximum light; cf.\ Fig.~\ref{optlowres}.

Molecular features that dominate the J band of S-type stars are identified by
\citet{Joy98} as due to TiO, VO, ZrO, ZrS, TiS, H$_2$O, CN, as well as
unidentified features, whereas C-type stars show only bands of the CN red 0-0
system. Positions of bands typical for the S-type stars are indicated in
Fig.~\ref{Jband}, however, these bands appear to be absent from both spectra.
This also holds true for the original, non-convolved NOTCam spectrum shown in
the upper panel of Fig.~\ref{Jband}.

\begin{figure}
  \centering
  \includegraphics[width=\columnwidth,bb=77 367 561 796, clip]{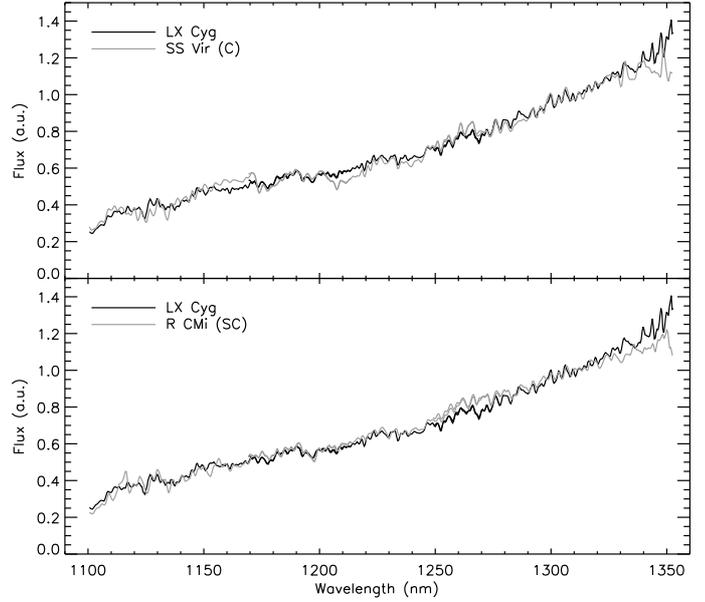}
  \caption{Comparison of the CRSP J-band spectrum of LX~Cyg to the carbon
    star SS~Vir (top) and the SC-type star R~CMi (bottom), obtained by
    \citet{Joy98} in 1994.}
  \label{SC-C}
\end{figure}

The CRSP spectrum of LX~Cyg was also compared to spectra from SC and carbon
stars obtained by \citet{Joy98}. A comparison with the C star \object{SS Vir}
and the SC star \object{R CMi} is shown in the upper and lower panel of
Fig.~\ref{SC-C}, respectively. The differences between the spectra are rather
subtle. There might be a feature around 1210\,nm in SS~Vir that is weaker or not
present in LX~Cyg and R~CMi. \citet{Joy98} mention that this feature could be
the R head of the C$_2$ Phillips $0-0$ band, but they did not find convincing
evidence for its presence in the higher-resolution spectra of the carbon stars. One
also needs to keep in mind the limited self-similarity of spectra of LX~Cyg
taken at different pulsation phases; cf.\ Fig.~\ref{Jband}. We conclude that the
J band alone is not suited to clearly distinguish between SC- and C-type stars
and that the exact spectral subtype of LX~Cyg in 1994 is unclear, although it was
not of type S.

\citet{Zij04} report observations of LX~Cyg in 2003, when it was claimed to have
been of spectral type SC. Unfortunately, no figure of that spectrum is shown and
the data are not available, so we are unable to compare to them.

Spectral type determinations in the 1960s and earlier report that bands of
LaO were present, and sometimes even strong. These bands seem to be absent in the
1970s, while ZrO was weakly present then. C$_2$ lines were weakly present in
1978. In particular,  all spectral type determinations up to 1975
are S, whereas they are SC from 1976 onwards. The observations available after
1979 are not conclusive enough to constrain the spectral type evolution. The new
data presented here suggest that LX~Cyg is a C-type star since 2008, when it
was observed by {\it Spitzer}. This means that the evolution from S to C type
was completed within 33 years between 1975 and 2008. Within roughly the same
period of time, the pulsation period increase happened simultaneously, strongly
suggesting a causal connection between these two phenomena.

%

\section{Discussion}\label{disc}

The observational evidence suggests that the pulsation period and spectral
type of LX~Cyg changed in parallel. A similar evolution has been previously
observed in the Mira BH~Cru \citep{TLE85,Whi99,Zij04,Utt11}, therefore, we may
refer to this as the BH~Cru phenomenon.

The hypothesis usually put forward to explain period changes of Mira stars is
that of an onset of a thermal pulse \citep{Wood75,WZ81}. However, the rate of
period change that one would expect from a TP is much lower than observed in a
BH~Cru phenomenon. Also, at the onset of a TP, the period is expected to
decrease because of the initial drop in luminosity caused by switching off the
H-burning shell. During no phase of a TP is such a strong period {increase} 
expected as is observed in LX~Cyg. Also, this scenario does not explain the
observed spectral evolution.

\citet{Zij04} discuss the evolution of BH~Cru and put forward another
hypothesis. They suggest that the spectral changes are caused by a decrease in
stellar temperature (by $\sim200$\,K), related to the increasing radius. They
also suggest that the evolution is unlikely to be related to an ongoing TP, for
similar arguments as above. Rather, they speculate that Mira periods may be
intrinsically unstable for ${\rm C}/{\rm O}\approx1$, possibly because of
feedback between the molecular opacities, pulsation amplitude, and period. Most
importantly, \citet{Zij04} refute that the C/O ratio changed in BH~Cru. From
this hypothesis one may however expect that some C-type stars can evolve back
to the S/SC type, which has never been observed. Also, the temperature
variability of a Mira may well exceed 200\,K during one pulsation cycle; for
instance, \citet{Loi01} find $\Delta T_{\rm eff}=400$\,K for BH~Cru. This should
induce the same spectral changes as the mechanism put forward by \citet{Zij04}
unless some of the involved molecular reaction rates are sufficiently slow so
that the feedback mechanism takes a few pulsation cycles.

Here, we follow a different hypothesis to explain the observed behaviour, which
was first proposed by \citet{Whi99}. We suggest that the period increase and
spectral evolution of LX~Cyg are the result of a 3DUP event that commenced in
the 1970s, which brought up an amount of $^{12}$C from the interior of the
star to its atmosphere. This raised the C/O ratio above unity and the star's
atmosphere became carbon rich. The genuine change in abundances led to a change
of the spectral appearance of the star because of the new molecular equilibrium
in the stellar atmosphere. At the same time as changing the spectral type, the
altered (enhanced) molecular opacity also influenced the stellar radius, which
led to an increase of the pulsation period on a much shorter timescale than
could be accomplished by the TP alone. The linear pulsation models presented in
\citet{LW07} indicate that indeed the pulsation period increases when the
chemistry changes from oxygen rich to carbon rich at a given luminosity.

The proposed scenario suggests that the spectral type and pulsation period
change are causally connected by the 3DUP event. It naturally explains both the
spectral changes and the period increase observed in BH~Cru and LX~Cyg and
avoids the problems of the alternative scenarios discussed above. According to
current evolutionary models of the TP-AGB, the TP preceding the 3DUP event must
have happened several hundred years ago
\citep[][; Cristallo, private communication]{Her05}. The time span of visual and
photometric observations for LX~Cyg is not long enough to identify the slow
period decrease expected after the onset of the TP on top of the fast evolution
induced by the 3DUP event.

A statistical argument may speak against the 3DUP hypothesis. Not many SC-type
stars are known, and out of this small sample two stars must be explained in
this way when the expected time span between two TPs (and 3DUP) may be 50\,000
to 70\,000 years for a plausible mass of 1.5\,$M_{\sun}$ \citep{VW93}. This would not be very likely to observe. On the other hand, a similarly rare event is that of
a very late thermal pulse, of which two, possibly even three examples are known
\citep{MB11}.

The upper limit of 33 years on the time it took for the spectral type change may
be compared to theoretical calculations of the time it takes to complete a full 3DUP
event. The timescale for this is governed by the velocity with
which the lower boundary of the convective envelope advances towards the core of
the star. Once the intershell material is swept up by the convective envelope,
it can be assumed to be mixed throughout the convective envelope very quickly
(i.e.\ within a few years) because the velocity of the convective mixing is very
fast, a few km\,s$^{-1}$. In the 3\,$M_{\sun}$ model considered by \citet{Mow99},
it takes $\sim120$ years for the surface carbon abundance to reach its
asymptotic value after the 3DUP; see their Fig.~7. However, the change in C
surface abundance by one 3DUP event may be much larger than that required for
an S/SC star with C/O very close to unity to convert it to a carbon star. Hence,
our observations do not necessarily  contradict these theoretical
considerations. This could also mean that the C surface abundance of LX~Cyg is
still on the rise and the 3DUP event still ongoing.

In principle, it would be desirable to measure the C/O ratio as a function of
time to elucidate the reason for the changes of period and spectral appearance
of the star. However, there are at least three obstacles that prevent us from
doing so:

\begin{enumerate}
\item Limitation by the data. Historical data are either not available or the
  spectral resolution and observed features do not allow for a reliable
  abundance determination.
\item Limitation by the variability of the star. LX~Cyg is a Mira variable with
  a very complex atmosphere and spectrum caused by the strongly pulsating
  atmosphere. Hydrostatic models are not suited for its analysis.
\item Limitation by the models. Hydrodynamic model atmosphere such as those from
  \citet{Hoe03} are available only for a few parameter combinations. These
  models of pulsating, carbon-rich giants are not intended, optimised, or tested
  for abundance determinations.
\end{enumerate}

We tried to constrain the present-day C/O ratio of LX~Cyg from our spectra,
however, it was impossible to achieve satisfying fits to any of the C/O-sensitive
spectral features with the help of hydrostatic model atmospheres. Estimates of
C/O ratios for SC-type stars were given in some of the older papers.
\citet{Ake79} measured a C/O abundance index of 5 and 6 from his two
observations of LX~Cyg, respectively. On his abundance scale, this corresponds
to ${\rm C/O}>0.95$ and $\sim1$ (his Table~4). On the other hand, \citet{KB80}
assign a C/O abundance index of $8-9$, which corresponds to a C/O ratio of
$1.00 - 1.02$ (Table~\ref{SC_criteria}).

\section{Conclusions}\label{conclusio}

We present spectral observations of the Mira variable LX~Cyg that cover a wide
range in wavelength and several years in time. We also discuss period
determinations of the star from the literature and complement this with an
analysis of recent AAVSO observations. Our own spectra and spectra available
from the literature are analysed to follow the spectral type change of LX~Cyg.
From this we conclude that:

The pulsation period of LX~Cyg was stable at $\sim460$\,d until 1975 when a
marked period increase of nearly 25\% commenced that lasted up to 1995, after
which it stabilised at $\sim580$\,d.
In parallel to the period increase, the spectral type changed from S via SC to
C. The available data do not well constrain when the spectral type change
happened, but a {\it Spitzer} IRS spectrum from 2008 suggests that LX Cyg was a
carbon star at that time.
Our spectral monitoring reveals that the star is of spectral type C during a
whole pulsation cycle.
The carbon star nature of LX~Cyg is corroborated by detailed comparisons with SC
and C star spectra available in the literature.
The {\it Spitzer} IRS spectrum suggests that SiC grains are present in the
circumstellar environment of LX~Cyg, possibly also PAHs. Also, the spectral
energy distribution and IR colours suggest that dust is present around the star.

The scenario of a recent 3DUP event presents a natural explanation for both the
spectral change and the period increase by the altered molecular equilibrium and
opacity by adding $^{12}$C to the atmosphere of the star. Both a recent thermal
pulse as well as a simple decrease in surface temperature do not satisfactorily
explain the observations.
Finally, we conclude that LX~Cyg is a rare case of a star where we can see AGB
stellar evolution in real-time and that the study of its general properties and
evolution may lead to unique insights into the transition from oxygen rich to
carbon rich stars on the AGB, which deserves further observations.

\begin{acknowledgements}
  The authors dedicate this paper to Tom Lloyd Evans who contributed to this
  work by private communications and discussions before he passed away
  unexpectedly on 12 June 2014. We thank Robin Lombaert for obtaining the
  Hermes/Mercator spectrum of LX~Cyg. This work would not have been possible
  without funding of SU by the Fund for Scientific Research of Flanders (FWO)
  under grant number G.0470.07 and by the Austrian Science Fund (FWF) under
  project P~22911-N16. TL has been supported by the Austrian Science Fund (FWF)
  under project number P~23737-N16. BA acknowledges the support from the
  {\em project STARKEY} funded by the ERC Consolidator Grant, G.A.\ n.~615604.
  LGR is co-funded under the Marie Curie Actions of the European Commission
  (FP7-COFUND). We thank the anonymous referee for helpful comments that
  stimulated improvements of the paper. We acknowledge with thanks the variable
  star observations from the AAVSO International Database contributed by
  observers worldwide and used in this research. This publication makes use of
  data products from the Two Micron All Sky Survey, which is a joint project of
  the University of Massachusetts and the Infrared Processing and Analysis
  Center/California Institute of Technology, funded by the National Aeronautics
  and Space Administration and the National Science Foundation.
\end{acknowledgements}

%
%

\end{document}